\documentclass[12pt]{article}

\usepackage{a4wide}
\usepackage{amsthm,amsfonts,amsmath}

\newcommand{\R}{\mathbb{R}} 
\newcommand{\C}{\mathbb{C}} 
\newcommand{\N}{\mathbb{N}} 
\newcommand{\sdots}{\, \cdot \,} 
\newtheorem{theorem}{Theorem}
\newtheorem{definition}{Definition}
\newtheorem{corollary}{Corollary}
\newtheorem{proposition}{Proposition}

\newtheorem{lemma}{Lemma}
\newcommand{\nor}[1]{\left\| #1 \right\|}
\newcommand{\caln}[1]{\nor{#1}} 
\newcommand{\calp}[1]{\nor{#1}_{\scriptscriptstyle \mathrm{A}}} 
\newcommand{\calq}[1]{\nor{#1}_{\scriptscriptstyle \mathrm{B}}}
\newcommand{\lgr}[1]{\left| #1 \right|}

\newcommand{\nocc}[2]{\left| #2 \right|_{#1}}
\newcommand{\seg}[2]{\left[ #1, #2 \right]}
\newcommand{\defeq}{\mathrel{\mathop:}=}
\newcommand{\ie}{\emph{i.e}}

\newcommand{\Nor}[1]{\nor{#1}_1}
\newcommand{\norinf}[1]{\nor{#1}_\infty}

\begin{document}



\sloppy 

\title{Asymptotic behavior of growth functions of D0L-systems}
\author{Julien Cassaigne \and Christian Mauduit \and Fran\c{c}ois Nicolas} 
\maketitle

\begin{abstract}
A \emph{D0L-system} is  a triple $(A, \sigma, w)$ where 
$A$ is a finite alphabet, 
$\sigma$ is an endomorphism of the free monoid over $A$, and
$w$ is a word over $A$.
The \emph{D0L-sequence} generated by  $(A, \sigma, w)$ is the sequence of words   $(w, \sigma(w), \sigma(\sigma(w)), \sigma(\sigma(\sigma(w))), \dotsc)$.
The corresponding sequence of lengths, \ie, the function mapping each integer $n \ge 0$ to $\lgr{ \sigma^n(w) }$, 
is called the \emph{growth function} of  $(A, \sigma, w)$.
In 1978,
Salomaa and Soittola 
deduced the following result from their thorough study of the theory of rational power series:
if the D0L-sequence generated by $(A, \sigma, w)$ is not eventually the empty word then there exist an integer $\alpha \ge 0$ and a real number $\beta \ge 1$
such that
$\lgr{\sigma^n(w)}$ behaves like $n^\alpha \beta^n$ as $n$ tends to infinity.
The aim of the present paper is to present a short, direct, elementary proof of this theorem.
\end{abstract}

\section{Introduction}

\subsection{Notation}

As usual,
 $\N$, $\R$ and $\C$ denote 
the semiring of natural integers, 
the field of real numbers,
and the field of complex numbers, respectively.
For every $a$, $b \in \N$, 
$\seg{a}{b}$ denotes the set of all integers $n$ such that $a \le n \le b$.
Let $f$, $g : \N \to \C$.
We write $f(n) \preceq g(n)$  if 
there exists a real number $\lambda > 0$ such that $\left\{ n \in \N : | f(n) | >  \lambda | g(n) | \right\}$ is finite.
We write $f(n) \asymp g(n)$  if both $f(n) \preceq g(n)$ and $g(n) \preceq f(n)$ hold.

A \emph{word} is a finite string of symbols. 
Word concatenation is denoted multiplicatively.
For every word $w$, the \emph{length} of $w$ is denoted  $\lgr{w}$. 
The word of length zero is called the \emph{empty word}.
For every symbol $a$ and every word $w$,
 $\nocc{a}{w}$ denotes the number of occurrences of $a$ in $w$.

An \emph{alphabet} is a finite set of symbols.
Let $A$ be an alphabet. 
The set of all words over $A$ is denoted $A^\star$.
A mapping $\sigma : A^\star \to A^\star$ is called a \emph{morphism} if $\sigma(xy) = \sigma(x) \sigma(y)$  for every $x$, $y \in A^\star$.
Clearly, 
$\sigma$ is completely determined  by its restriction to $A$.
For every $n \in \N$, $\sigma^n$ denotes the $n^\text{th}$ iterate of $\sigma$: 
for every $w \in A^\star$, 
$\sigma^0(w) = w$, 
$\sigma^1(w) = \sigma(w)$,
$\sigma^2(w) = \sigma(\sigma(w))$, 
$\sigma^3(w) = \sigma(\sigma(\sigma(w)))$, 
etc.

A \emph{D0L-system} \cite{Lindenmayer68} is defined as a triple $(A, \sigma, w)$ where 
$A$ is an alphabet, 
$\sigma$ is a morphism from $A^\star$ to itself, 
and 
$w$ is a  word over $A$. 
The \emph{growth function} of the D0L-system $(A, \sigma, w)$ is defined as the integer sequence 
$\left(\lgr{w}, \lgr{\sigma(w)}, \lgr{\sigma^2(w)}, \lgr{\sigma^3(w)}, \dotsc \right)$.
For every  \emph{D0L-system} $(A, \sigma, w)$, either the  sequence  $(w, \sigma(w), \sigma^2(w), \sigma^3(w), \dotsc)$ is eventually periodic or $\lim_{n \to \infty} \lgr{\sigma^n(w)} = \infty$.

\subsection{Contribution}

The aim of the paper is to present a short, elementary proof of the following theorem.

\begin{theorem}
\label{th:D0L-growth} 
Let $(A, \sigma, w)$ be a D0L-system such that $\sigma^n(w)$ is a non-empty word for every $n \in \N$.
There exist a non-negative integer $\alpha$ smaller than the cardinality of $A$,
 and a real number $\beta \ge 1$ such that $\lgr{\sigma^n(w)} \asymp n^\alpha \beta^n$ as $n \to \infty$.
\end{theorem}

Theorem~\ref{th:D0L-growth} plays a crucial role in the proof of an important result: 
Pansiot's theorem concerning the complexity of pure morphic sequences \cite{Pansiot84}.

In 1978, Salomaa and Soittola laboriously proved a stronger result than Theorem~\ref{th:D0L-growth}.

\begin{theorem}[Salomaa and Soittola \cite{SalomaaS78,BerstelR88}] \label{th:SS}
Let $(A, \sigma, w)$ be  a D0L-system such that $\sigma^n(w)$ is a non-empty word for every $n \in \N$.
There exist 
a positive integer $q$,
a non-negative integer $\alpha$  smaller than the cardinality of $A$, 
 and
a real number $\beta \ge 1$ such that
for each $r \in \seg{0}{q - 1}$, 
$$
\frac{{\lgr{\sigma^{nq + r}(w)}}}
{{(n q + r)}^{\alpha} {\beta}^{nq + r} }
$$
converges to a positive, finite limit as $n \to \infty$.
\end{theorem}

The proof of Theorem~\ref{th:D0L-growth} presented below cannot likely be refined into a proof of Theorem~\ref{th:SS}. 
The original proof of  Theorem~\ref{th:SS} relies on the theory of rational power series.
In particular, two deep results are put to use:
 \begin{enumerate}
\item  Sch\"utzenberger's representation theorem  \cite{SalomaaS78,BerstelR88}, and
\item Berstel's theorem concerning the minimum-modulus poles of univariate rational  series over the semiring of non-negative real numbers  \cite{SalomaaS78,BerstelR88}.
\end{enumerate}

To conclude this section note  that a very  interesting particular case of Theorem~\ref{th:SS} can be simply deduced from the Perron-Frobenius theory.

\begin{definition}[Irreducibility and  period]
Let $A$ be an alphabet and let $\sigma : A^\star \to A^\star$ be a  morphism.
We say that $\sigma$ is \emph{irreducible} if
 for each $(a, b) \in A \times A$, 
 there exists $k \in \N$ such that $a$ occurs in $\sigma^k(b)$.  
For every $a \in A$, the \emph{period} of $a$ under $\sigma$ is defined as the greatest common divisor of $\left\{ k \in \N : \nocc{a}{\sigma^k(a)} \ne 0 \right\}$.
\end{definition}

If the morphism $\sigma$ is irreducible then all letters in $A$ have the same period under $\sigma$.
If  $\sigma$ is irreducible and  if every letter in $A$ is of period one under $\sigma$ then  $\sigma$ is  called \emph{primitive}: there exists $N \in \N$ such that  for each $(a, b) \in A \times A$, 
$a$ occurs in $\sigma^N(b)$.

\begin{theorem}[\cite{Queffelec87}] \label{th:Queffelec}
Let $A$ be an alphabet, let $\sigma : A^\star \to A^\star$ be an irreducible morphism, and let $q$ denote the period under $\sigma$ of any letter in $A$.
There exists a real number $\beta \ge 1$ such that for each $(a, b) \in A \times A$ and each $r \in \seg{0}{q - 1}$,
$$
\frac{\nocc{b}{\sigma^{nq + r}(a)}}{ {\beta}^{nq + r} }
$$
converges to a positive, finite limit as $n \to \infty$.
\end{theorem}

\section{Proof of  Theorem~\ref{th:D0L-growth}} \label{sec:proof}

Our proof of Theorem~\ref{th:D0L-growth} relies on the equivalence of norms on a finite-dimensional vector space (see Theorem~\ref{th:equiv-norm} below).
For the sake of completeness, the definition of a norm is recalled.

\begin{definition}[Norm]
Let $V$ be a real or complex vector space.
A \emph{norm} on $V$ is a mapping $\caln{\sdots}$ from $V$ to $\R$ such that the following three properties hold for all vectors $x$, $y \in V$ and all scalars $\lambda \in \R$:
\begin{enumerate}
\item  $\caln{x} = 0$ if, and only if, $x$ is the zero vector,
\item $\caln{\lambda x} = | \lambda |  \caln{x}$, and 
\item $\caln{x +  y} \le  \caln{x} + \caln{y}$.
\end{enumerate}
\end{definition}

\begin{theorem}[{\cite[Corollary~3.14]{Lang83}}] \label{th:equiv-norm}
Let $V$ be a real or complex  vector space.
If the dimension of $V$ is finite then for any norms $\calp{\sdots}$ and $\calq{\sdots}$ on $V$, 
there exist  positive real numbers $\lambda$ and $\mu$ such that $\lambda \calp{x} \le \calq{x} \le  \mu \calp{x}$ for every $x \in V$.
\end{theorem}

Throughout this section $d$ denotes a positive integer and $\C^{d \times d}$ denotes the algebra of $d$-by-$d$ complex matrices.
The following two classical norms on $\C^{d \times d}$ play a central role in our discussion.

\begin{definition}
For every $X \in \C^{d \times d}$,
define $\Nor{X}$ as the \emph{Manhattan norm} of $X$: 
$\Nor{X}$  equals the sum of the magnitudes of the entries of $X$. 
\end{definition}

\begin{definition}
For every $X \in \C^{d \times d}$,
define $\norinf{X}$ as the \emph{maximum norm} of $X$: 
$\norinf{X}$  equals the maximum magnitude of the entries of $X$.
\end{definition} 

It is clear that $\norinf{X} \le  \Nor{X} \le d^2 \norinf{X}$ for every $X \in \C^{d \times d}$.

The next proposition, which is mainly folklore, 
is the main ingredient of the proof of Theorem~\ref{th:D0L-growth}.

\begin{proposition} \label{prop:norm-mat}
For each non-nilpotent matrix $M \in \C^{d \times d}$,
there exist 
a norm  $\caln{\sdots}$ on $\C^{d \times d}$,
an integer $\alpha \in \seg{0}{d - 1}$
and  a real number $\beta >  0$  such that the ratio
$
\dfrac{\caln{M^n}}{n^\alpha \beta^n} 
$
converges to a positive, finite limit as $n \to \infty$.
\end{proposition} 

\begin{proof}
Let $P  \in \C^{d \times d}$ be a non-singular matrix such that 
$P M P^{-1}$ is in Jordan normal form:
there exist $D$, $N \in \C^{d \times d}$ 
such that 
$D$ is diagonal, 
$N$ is nilpotent,   
$P M P^{-1} = D + N$
and 
$D N = N D$.
Let $\caln{\sdots}$ be the norm on $\C^{d \times d}$ defined by: $ \caln{X} \defeq \norinf{P X P^{-1}}$ for every $X \in \C^{d \times d}$.

For all $i$, $j \in \seg{1}{d}$, 
let $e_{i, j} : \N \to \C$ be the function mapping each $n \in \N$ to the  ${(i, j)}^\text{th}$ entry of $P M^n P^{-1}$.
It is clear that $\caln{M^n} = \max_{i, j \in \seg{1}{d}} | e_{i, j}(n) |$  for every $n \in \N$.
Let $I$ be the set of all $(i, j) \in \seg{1}{d} \times \seg{1}{d}$ such that $e_{i, j} $ is not eventually zero.
Since $M$ is not nilpotent, $I$ is non-empty, 
and thus 
$$
\caln{M^n} = \max_{(i, j) \in I} | e_{i, j}(n) | 
$$
for every sufficiently large $n \in \N$.

For each  $n \in \N$, 
the binomial theorem yields:
$$
P M^n P^{-1} 
= 
{(D + N)}^{n}
= 
\sum_{k = 0}^n \dbinom{n}{k} D^{n - k} N^{k}  \, .
$$
Besides, 
$N^k$ is a zero matrix for every integer $k \ge d$, 
and thus 
$$
P M^n P^{-1}  = \sum_{k = 0}^{d - 1} \dbinom{n}{k} D^{n - k} N^{k} 
$$
for every integer $n \ge d - 1$.
Hence, for each $(i, j) \in I$, 
there exist a non-zero eigenvalue $\lambda_i$ of $D$ and a non-zero complex polynomial  $f_{i, j}$ with $\deg f_{i, j} \le d - 1$ 
such that 
$$
e_{i, j}(n) = f_{i, j}(n) \lambda_i^n 
$$
for every integer $n \ge d - 1$:

Let $(\beta, \alpha)$ be the maximum element of $\left\{ \left(| \lambda_i |, \deg f_{i, j} \right) : (i, j) \in I \right\}$
according to the lexicographical order.
Let $J$ denote the set of all $(i, j) \in I$ such that $ \left(| \lambda_i |, \deg f_{i, j} \right) = (\beta, \alpha)$, and 
for each $(i, j) \in J$, let $c_{i, j}$ denote the leading coefficient of $f_{i, j}$.
It is clear that 
$$
\lim_{n \to \infty} \frac{| e_{i, j}(n) | }{n^\alpha \beta^n} =  
\begin{cases}
| c_{i, j} | & \text{if $(i, j) \in J$} \\
0 & \text{otherwise}  
\end{cases}
$$
for every $(i, j) \in I$, so 
$$
\lim_{n \to \infty} \dfrac{\caln{M^n}}{n^\alpha \beta^n}  
= 
\max_{(i, j) \in I} \left( \lim_{n \to \infty}  \frac{| e_{i, j}(n) | }{n^\alpha \beta^n}  \right)
=
\max_{(i, j) \in J} | c_{i, j} | \, .$$
\end{proof}

It follows from Theorem~\ref{th:equiv-norm}
that for any norms  $\calp{\sdots}$ and $\calq{\sdots}$ on $\C^{d \times d}$
and  
for any $M \in \C^{d \times d}$, 
$\calp{M^n} \asymp \calq{M^n}$ as $n \to \infty$, so we get:

\begin{corollary} \label{cor:equiv-norm}
For each matrix $M \in \C^{d \times d}$,
there exist 
an integer $\alpha \in \seg{0}{d - 1}$
and  a real number $\beta \ge 0$  such that 
for every norm $\caln{\sdots}$ on $\C^{d \times d}$,
$\caln{M^n} \asymp n^\alpha \beta^n$ as $n \to \infty$.
\end{corollary}

Proposition~\ref{prop:norm-mat} deserves several comments.
First, a more precise result is known.

\begin{theorem}[{\cite[Theorem~3.1]{Varga1962}}] \label{th:Varga}
Let $\nor{\sdots}$ denote the spectral norm on $\C^{d \times d}$
and 
let $M \in \C^{d \times d}$ be such that $M$ is not nilpotent.
\begin{itemize}
\item 
Let $\beta$ denote the spectral radius of $M$. 
\item 
Let $j$ denote the maximum size of the Jordan blocks of $M$ with spectral radius $\beta$.
\end{itemize}
The ratio $\dfrac{\nor{M^n}}{n^{j - 1} \beta^n}$ converges to a positive, finite limit as $n \to \infty$.
\end{theorem}

Let us also mention that a weak version of 
Theorem~\ref{th:Varga}
holds in an arbitrary Banach algebra.   

\begin{theorem}[Gelfand's formula \cite{Rudin3}] 
Let $\mathcal{A}$ be a complex Banach algebra and let $\nor{\sdots}$ denote its norm.
For every $M \in \mathcal{A}$, $\sqrt[n]{\nor{M^n}}$ converges to the spectral radius of $M$ as $n \to \infty$.
\end{theorem}

Let us now illustrate Proposition~\ref{prop:norm-mat} and Corollary~\ref{cor:equiv-norm} with an example.
The matrix 
$$
M \defeq 
\begin{bmatrix} 
4 & -3 \\
3 & 4 
\end{bmatrix}
$$
is diagonalizable:
$$
P M P^{-1} = 
\begin{bmatrix} 
\lambda & 0 \\
0 & \bar \lambda 
\end{bmatrix} \, ,
$$
where $i$ denotes the imaginary unit,
\begin{align*}
\lambda & \defeq 4 + 3i\,, 
&
\bar \lambda & \defeq 4 - 3i\,, 
&
P & \defeq 
\begin{bmatrix} 
1 & i \\
1 & - i 
\end{bmatrix}
& \text{and} & & 
P^{-1} & \defeq \dfrac{1}{2} 
\begin{bmatrix} 
1 & 1 \\
-i & i 
\end{bmatrix}  \, .
\end{align*}
Let $\caln{\sdots}$ be the norm on $\C^{2 \times 2}$ defined by: 
$\caln{X} \defeq \norinf{P X P^{-1}}$ for every $X \in \C^{2 \times 2}$.
For every $n \in \N$, we have
$$
M^n 
 = P^{-1} \begin{bmatrix} 
\lambda^n & 0 \\
0 & \bar \lambda^n 
\end{bmatrix} P 
=
\frac{1}{2} 
\begin{bmatrix} 
\lambda^n + \bar \lambda^n &  i \lambda^n - i \bar \lambda^n  \\
- i \lambda^n + i  \bar \lambda^n  & \lambda^n + \bar \lambda^n 
\end{bmatrix}
= 
5^n
\begin{bmatrix} 
\cos(n \theta) &  -\sin(n \theta)  \\
\sin(n \theta) &  \cos(n \theta)
\end{bmatrix} \, , 
$$
where $\theta$ is an argument of $\lambda$; so 
\begin{align*}
\frac{ \nor{M^n}}{5^n} & = 1 \,,
&
2 & \le  \frac{\Nor{M^n}}{5^n}   \le 2 \sqrt{2} 
& \text{and} & &   
\frac{\sqrt{2}}{2}  & \le \frac{\norinf{M^n}}{5^n} \le  1
 \, .
\end{align*}
Noteworthy is that no entry of $5^{-n}M^n$ converges as $n \to \infty$:
both sets 
$\left\{ \cos(n \theta) : n \in \N \right\}$
and  
$\left\{ \sin(n \theta)  : n \in \N \right\}$
are  dense subsets of  the closed real interval with endpoints $-1$ and $+1$
(see appendix).

We turn back to the proof of Theorem~\ref{th:D0L-growth}.

\begin{lemma} \label{lem:occur-O}
Let $A$ be an alphabet, let $\sigma : A^\star \to A^\star$ be a morphism and let $w$, $x \in A^\star$. 
If  $x$ occurs in $\sigma^{n_0}(w)$ for some $n_0 \in \N$ 
then 
$\lgr{\sigma^n(x)} \preceq \lgr{\sigma^n(w)}$ as $n \to \infty$.
\end{lemma}

\begin{proof}
Let $L \defeq \max_{a \in A} \lgr{\sigma^{n_0}(a)}$.
If $x$ occurs in $\sigma^{n_0}(w)$ then for every $n \in \N$, 
$\sigma^n(x)$ occurs in $\sigma^{n + n_0}(w)$,
 and thus 
$$\lgr{\sigma^n(x)} \le \lgr{\sigma^{n + n_0}(w)}
= \sum_{a \in A} \nocc{a}{\sigma^n(w)}  \lgr{\sigma^{n_0}(a)} 
  \le L \sum_{a \in A} \nocc{a}{\sigma^n(w)} 
 = L \lgr{\sigma^n(w)}
 \, .
$$
\end{proof}

\begin{definition}
A D0L-system  $(A, \sigma, w)$  is called \emph{reduced} if for every $a \in A$ there exists $m \in \N$ such that $a$ occurs in $\sigma^m(w)$. 
\end{definition}


\begin{lemma} \label{lem:asymp-sum}
For any reduced D0L-system  $(A, \sigma, w)$, 
\begin{equation} \label{eq:sigmaw-suma}
 \lgr{\sigma^n(w)} \asymp 
\sum_{a \in A} \lgr{\sigma^n(a)}     
\end{equation}
as $n \to \infty$.
\end{lemma}

\begin{proof}
For every $n \in \N$, let 
$$
S_n \defeq \sum_{a \in A} \lgr{\sigma^n(a)} \, .
$$
First, we have 
$$
\lgr{\sigma^n(w)} 
=  \sum_{a \in A} \nocc{a}{w} \lgr{\sigma^n(a)} 
\le 
\left( \max_{a \in A} \nocc{a}{w} \right)
S_n  \, , 
$$
and thus $\lgr{\sigma^n(w)}  \preceq S_n$.
Conversely, Lemma~\ref{lem:occur-O} ensures $\lgr{\sigma^n(a)}  \preceq \lgr{\sigma^n(w)}$ for each $a \in A$ 
because the D0L-system $(A, \sigma, w)$ is reduced.
It follows $S_n \preceq  \lgr{\sigma^n(w)}$.
\end{proof}

\begin{proof}[Proof of Theorem~\ref{th:D0L-growth}]
Let us first check that, without loss of generality, we  may assume that $(A, \sigma, w)$ is reduced.
Let $\bar A$ denote the set of all symbols $a \in A$ such that $a$ occurs in $\sigma^m(w)$ for some $m \in \N$.
Remark that $\sigma(\bar A) \subseteq \bar A^\star$: 
for any $a \in \bar A$ and any $m \in \N$ such that $a$ occurs in $\sigma^m(w)$,
 $\sigma(a)$ occurs in $\sigma^{m + 1}(w)$, and thus $\sigma(a) \in \bar A^\star$. 
Hence $\sigma$ induces a morphism $\bar \sigma : \bar A^\star \to \bar A^\star$: 
 $\bar \sigma(x) =   \sigma(x)$ for every $x \in \bar A^\star$.
Clearly, $(\bar A, \bar \sigma, w)$ is a reduced D0L-system and   $\sigma^n(w) = \bar \sigma^n(w)$ for every  $n \in \N$.
Therefore, we may replace $(A, \sigma, w)$ with $(\bar A, \bar \sigma, w)$ in the remaining of the proof, so  \eqref{eq:sigmaw-suma} holds by Lemma~\ref{lem:asymp-sum}.

Let $d$ denote the cardinality of $A$.
Write arbitrarily $A$ in the form  $A = \{ a_1, a_2, \dotsc, a_d \}$.
Let $M$ be the $d$-by-$d$ matrix defined by: 
 for all $i$, $j \in \seg{1}{d}$, the ${(i, j)}^\text{th}$ entry of $M$ equals $\nocc{a_i}{\sigma(a_j)}$.
The ${(i, j)}^\text{th}$ entry of $M^n$ equals $\nocc{a_i}{\sigma^n(a_j)}$,
 and thus
\begin{equation} \label{eq:sum-norm}
\sum_{a \in A} \lgr{\sigma^n(a)} = \Nor{M^n}		
\end{equation} 

It follows from Corollary~\ref{cor:equiv-norm} that there exist 
an integer $\alpha \in \seg{0}{d - 1}$
and  a real number $\beta \ge  0$  such that
\begin{equation} \label{eq:NorMn-alpha-beta}
 \Nor{M^n}  \asymp n^{\alpha} \beta^n		
\end{equation}%

Combining \eqref{eq:sigmaw-suma},  \eqref{eq:sum-norm}  and~\eqref{eq:NorMn-alpha-beta}, 
we get $\lgr{\sigma^n(w)} \asymp n^{\alpha} \beta^n$. 
Since $\lgr{\sigma^n(w)} \ge 1$ for every $n \in \N$, 
$n^{\alpha} \beta^n$ does not converge to zero, 
and thus $\beta \ge 1$.
\end{proof}

\bibliographystyle{plain}
\bibliography{BibPansiot}

\begin{thebibliography}{10}

\bibitem{BerstelR88}
J.~Berstel and C.~Reutenauer.
\newblock {\em Rational series and their languages}, volume~12 of {\em EATCS
  Monographs on Theoretical Computer Science}.
\newblock Springer-Verlag, 1988.
\newblock The new version is presently available online at Berstel's homepage.

\bibitem{HardyWright}
G.~H. Hardy and E.~M. Wright.
\newblock {\em An introduction to the theory of numbers}.
\newblock Oxford, at the Clarendon Press, fourth edition, 1979.

\bibitem{KaplanskyGreen}
I.~Kaplansky.
\newblock {\em Commutative rings}.
\newblock The University of Chicago Press, revised edition, 1974.

\bibitem{Lang83}
S.~Lang.
\newblock {\em Real analysis}.
\newblock Addison-Wesley Publishing Company, second edition, 1983.

\bibitem{Lindenmayer68}
A.~Lindenmayer.
\newblock Mathematical models for cellular interactions in development.
\newblock {\em Journal of Theoretical Biology}, 18(3):280--315, 1968.

\bibitem{NivenZuck}
I.~Niven and H.~S. Zuckerman.
\newblock {\em An introduction to the theory of numbers}.
\newblock John Wiley and Sons, third edition, 1972.

\bibitem{Pansiot84}
J.-J. Pansiot.
\newblock Complexit{\'e} des facteurs des mots infinis engendr{\'e}s par
  morphismes it{\'e}r{\'e}s.
\newblock In {\em Proceedings of the 11th International Colloquium on Automata,
  Languages and Programming (ICALP'84)}, volume 172 of {\em Lecture Notes in
  Computer Science}, pages 380--389. Springer-Verlag, 1984.

\bibitem{Queffelec87}
M.~Queff\'elec.
\newblock {\em Substitution dynamical systems-spectral analysis}, volume 1294
  of {\em Lecture Notes in Mathematics}.
\newblock Springer-Verlag, 1987.

\bibitem{Rudin3}
W.~Rudin.
\newblock {\em Functional analysis}.
\newblock International Series in Pure and Applied Mathematics. McGraw-Hill,
  second edition, 1991.

\bibitem{SalomaaS78}
A.~Salomaa and M.~Soittola.
\newblock {\em Automata-theoretic aspects of formal power series}.
\newblock Texts and Monographs in Computer Science. Springer-Verlag, 1978.

\bibitem{Varga1962}
R.~S. Varga.
\newblock {\em Matrix iterative analysis}.
\newblock Prentice-Hall, 1962.

\end{thebibliography}

\appendix

\section*{Appendix} \label{app:trigo}

Throughout the  section, 
\begin{itemize}
\item $\pi$ denotes Archimedes' constant,  
\item $I \defeq \left\{ x \in \R : 0 \le x \le 1 \right\}$, and
\item  $J\defeq \left\{ x \in \R : - 1 \le x \le +1 \right\}$.
\end{itemize}
The aim of this appendix is to prove the following proposition:

\begin{proposition} \label{prop:dense-theta}
For any argument $\theta$ of  $4 + 3i$, both sets 
$\left\{ \cos(n \theta) : n \in \N \right\}$
and  
$\left\{ \sin(n \theta)  : n \in \N \right\}$
are dense subsets  of $J$.
\end{proposition}

Proposition~\ref{prop:dense-theta} is a consequence of the following two well-known results.

\begin{proposition}[{\cite[Theorem~6.15]{NivenZuck}}]  \label{prop:cos-irrational}
Let $\rho$ be a rational number.
If $\cos(2 \pi  \rho)$ is rational then $2 \cos(2 \pi  \rho)$ is an integer.
\end{proposition} 

\begin{proof}
Both complex numbers $\exp( 2 \pi \rho i)$ and  $\exp(- 2 \pi \rho i)$ are algebraic integers.
Indeed, they are roots of the monic integer polynomial $z^q - 1$, 
where $q$ is a positive integer such that $q \rho$ is an integer.
Since a sum of algebraic integers is also an algebraic integer \cite[Theorem~13]{KaplanskyGreen}, 
$2 \cos (2 \pi  \rho) =  \exp( 2 \pi \rho i) + \exp(- 2 \pi \rho i)$ is an algebraic integer.
If  $ \cos (2 \pi  \rho)$ is rational then $2 \cos (2 \pi  \rho)$ is in fact an integer 
because an algebraic integer, if rational, is an integer \cite[Theorem~206]{HardyWright}.
\end{proof} 

Note that for any real number $\theta$ with $- \pi \le \theta \le \pi$, 
the following three  assertions are equivalent:
\begin{enumerate}
 \item 
 $2 \cos(\theta)$  is an integer, 
 \item $\cos(\theta) \in \left\{ - 1, - \frac{1}{2}, 0, + \frac{1}{2}, + 1 \right\}$, and
 \item 
$|\theta| \in \left\{  0, \frac{1}{3}\pi, \frac{1}{2} \pi, \frac{2}{3} \pi, \pi \right\}$.
\end{enumerate}

\begin{proposition}[{\cite[Theorem~439]{HardyWright}}] \label{prop:rho-irrational}
For any irrational number $\rho \in \R$,
 $\left\{  n \rho  - \lfloor{n \rho} \rfloor : n \in  \N \right\}$  is a dense subset of $I$.
\end{proposition}

\begin{proof}[Proof of Proposition~\ref{prop:dense-theta}]
Since the cosine of $\theta$ equals $\frac{4}{5}$, 
$\frac{\theta}{ 2 \pi }$ is irrational by Proposition~\ref{prop:cos-irrational}.
Hence, 
$ D \defeq \left\{  \frac{ n \theta}{ 2 \pi }  - \left\lfloor { \frac{n \theta}{ 2 \pi }} \right\rfloor 
  : n \in  \N  \right\}$
is a dense subset of $I$ by Proposition~\ref{prop:rho-irrational}.
Since the function $f : I \to J$ that maps each $x \in I$ to  $\cos(2 \pi x)$   is continuous and surjective, $\left\{ \cos( n \theta)  : n \in \N \right\} = f(D)$ is a dense subset of $J$.
In the same way, the function $g : I \to J$ that maps each $x \in I$ to  $\sin(2 \pi x)$ is continuous, surjective and such that  $\left\{ \sin( n \theta)  : n \in \N \right\} = g(D)$.
 \end{proof}

\end{document}